\documentclass[journal=jpclck,manuscript=letter]{achemso}

\usepackage{chemformula} 
\usepackage[T1]{fontenc} 
\usepackage{dsfont}
\usepackage{hyperref}

\author{Noah Garrett}
\author{Michael Rose}
\author{David A. Mazziotti}
\affiliation{Department of Chemistry and The James Franck Institute, \\ The University of Chicago, Chicago, IL 60637 USA}
\email{damazz@uchicago.edu}

\title[An \textsf{achemso} demo]
  {Size-Consistent Quantum Chemistry on Quantum Computers}

\abbreviations{IR,NMR,UV}
\keywords{American Chemical Society, \LaTeX}

\begin{document}

\begin{tocentry}

\centering
\includegraphics[height=4.50cm]{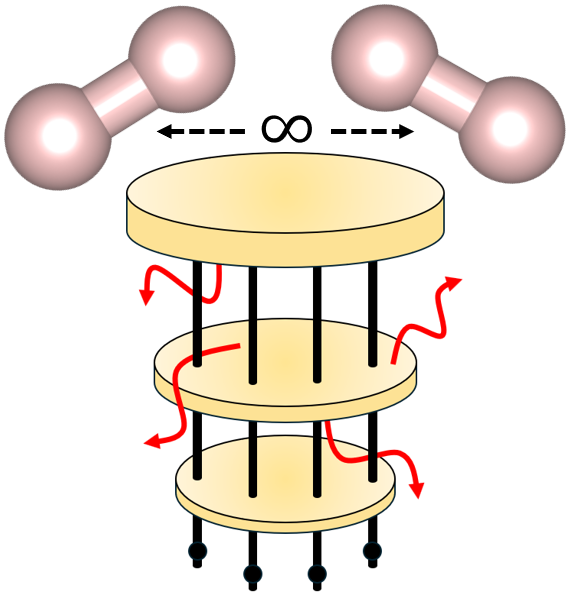}

\end{tocentry}

\begin{abstract}


Hybrid quantum–classical algorithms have begun to leverage quantum devices to efficiently represent many-electron wavefunctions, enabling early demonstrations of molecular simulations on real hardware. A key prerequisite for scalable quantum chemistry, however, is size consistency: the energy of non-interacting subsystems must scale linearly with system size. While many algorithms are theoretically size-consistent, noise on quantum devices may couple nominally independent subsystems and degrade this fundamental property. Here, we systematically evaluate size consistency on quantum hardware by simulating systems composed of increasing numbers of non-interacting \ch{H2} molecules using optimally shallow unitary circuits. We find that molecular energies remain size-consistent within chemical accuracy for an estimated 118 and 71 \ch{H2} subsystems for one- and two-qubit unitary designs, respectively, demonstrating that current quantum devices preserve size consistency over chemically relevant system sizes and supporting the feasibility of scalable, noise-resilient simulation of strongly correlated molecules and materials.

\end{abstract}


\newpage

When a molecular bond is stretched to the infinite limit, two non-interacting subsystems are created. As they can no longer influence each other, predicting observables of these subsystems independently or simultaneously should yield identical results; this agreement is called size consistency. Size consistency is a key principle of electronic structure methods to give energies and properties that are unchanged when a non-interacting subsystem is added at infinite separation.\cite{Pople1976, Bartlett1977, Barlett1981, Cremer1987, Deleuze1995, Sheppard1984, Szabo1996, Mahapatra1998, Mahapatra1999, Deleuze2003, Nooijen2005, Hirata2011, Sen2015, Mihalka2020, Xiao2024, Kong2024} Violations of size consistency lead to systematic errors that grow with system size, preventing meaningful comparison of thermodynamic and reaction energies across fragments. This requirement is therefore critical for modeling fragmented or non-interacting molecular systems found in bond dissociation and chemical reactions. Furthermore, size consistency plays an important role in the proper scaling of the energy with system size,\cite{Pople1976, Bartlett1977, Barlett1981, Cremer1987, Szabo1996} which is necessary for accurately modeling large, strongly correlated systems such as superconductors, quantum sensors, and other correlated materials.\cite{Bardeen1957, Dagotto1994, Bennemann2004, Lee2006, Giovannetti2004, Talor2008, Degen2017, Ostroverkhova2016, Dagotto2005} These systems are characterized by complex electron interactions and near-degenerate states, which require multi-reference or active space methods capable of describing multiple electron configurations.\cite{Mazziotti2007.3, Ganoe2024, Schouten2025} However, these methods scale combinatorially with the number of orbitals and electrons, making them infeasible on classical computers for large systems of interest.\cite{Roos1980, Roos1987, Szabo1996}

Since qubits naturally represent superposition and entanglement, quantum computers have the capability to store the entire wavefunction in a compact representation of qubits, avoiding the costly scaling of classical approaches and enabling the accurate study of strongly correlated systems.\cite{Feynman1982, McArdle2020, Motta2022, Smart2022.6, Smart2022.9, Wang2023.9, Wang2023.10} As a result, the demonstration of size consistency is a key prerequisite to performing large molecular simulations on quantum computers and to obtaining quantum advantage.\cite{Cao2019, Bauer2020, McArdle2020, Motta2022} While many methods themselves are designed to be size-consistent, the effect of noise on noisy intermediate-scale quantum (NISQ) devices must also be evaluated for size consistency.\cite{Preskill2018} The presence of noise on a quantum computer may alter the outcome of simulations in ways that disrupt the expected scaling behavior, potentially limiting the accuracy and reliability of calculations as system size increases.

In this Letter, we systematically assess the ability of quantum hardware to perform size-consistent molecular simulations, laying the foundation for scalable, strongly correlated many-body calculations. To directly evaluate the effect of noise, we employ a method-agnostic strategy using optimally shallow unitary transformations applied to systems composed of increasing numbers of non-interacting \ch{H2} molecules. By computing the energy and ground-state double-excitation population per \ch{H2} as a function of system size on quantum hardware, we show that observed energies remain within chemical accuracy for small subsystem sizes and that reliable scaling behavior persists as noise increases for larger numbers of subsystems. These results demonstrate that present-day devices can preserve size consistency across chemically relevant system sizes despite noise, providing a necessary foundation for scalable quantum simulations of strongly correlated molecules and materials.


Size consistency requires that the observables of non-interacting subsystems can be predicted equivalently, whether computed simultaneously or independently. This property implies two separability conditions for non-interacting systems: additive separability, which requires that the total energy equals the sum of the subsystems' energies, and multiplicative separability, which necessitates that the total wavefunction is the antisymmetrized product of each individual wavefunction.\cite{Nooijen2005,Pople1976} Truncated methods like configuration interaction with single and double excitations (CISD) fail to preserve these properties. Consider a supermolecule composed of two non-interacting \ch{H2} molecules, where each \ch{H2} ground state in a minimal basis can be modeled with two determinants corresponding to a mean-field reference and its double excitation.\cite{Szabo1996} The Hamiltonian of the total system is the sum of the Hamiltonian for each subsystem embedded in the larger Hilbert space
\begin{equation}
\hat{H}_{AB} = (\hat{H}_{A} \otimes \mathds{1}) + ( \mathds{1} \otimes \hat{H}_{B}).
\end{equation}
This form implies that the total energy should equal the sum of the energy for each molecule, $E_{AB} = E_A + E_B$. By contrast, treating this system with a CISD wavefunction does not include the quadruple excitation determinant that appears when each \ch{H2} contributes a double excitation to the ground state. Since the antisymmetrized product of the two subsystem wavefunctions contains these higher-order excitations, CISD cannot represent the correct product state and therefore violates multiplicative separability and is not size consistent. These missing configurations also prevent the CISD energies from scaling with the sum of the subsystems and fail to preserve additive separability.\cite{Nooijen2005} As the system size increases to $N$ non-interacting \ch{H2} molecules, the number of quadruple, sextuple, and higher-order excitations, not included by CISD, grows combinatorially, and the correlation energy $E_{\mathrm{Corr}}$ per \ch{H2}
\begin{equation}
    \lim_{N\to\infty} \frac{^NE_{\mathrm{Corr}}(\mathrm{CISD})}{N} = 0
\end{equation}
vanishes in the limit of infinite molecules.\cite{Szabo1996}

Since CISD omits increasingly important higher-order excitations needed to describe long-range correlation and collective behavior, size-inconsistent methods lead to significant errors that grow with systems size, making them infeasible for periodic systems,\cite{Deleuze1995,Szabo1996,Deleuze2003,Hirata2011} such as lattice structures or layered materials.\cite{Makov1995,Kratzer2019} This behavior reflects a breakdown of size-extensivity, which requires proper scaling of the energy with respect to system size or number of electrons.\cite{Barlett1981,Mahapatra1998, Mahapatra1999, Nooijen2005} For closed-shell, non-interacting systems, size consistency ensures linear scaling and can be viewed as a requirement for size extensivity.\cite{Nooijen2005} In addition, violation of size extensivity often coincide with failures of size consistency. Furthermore, in a chemical reaction such as

\begin{equation}
{\rm N}_2 + 3 {\rm H}_2 \rightarrow 2 {\rm NH}_3
\end{equation}

\noindent accurately determining the reaction energy $\Delta E$ requires computational methods that can consistently model the non-interacting fragments \ch{N2} and \ch{H2} independently and the compound system \ch{NH3}.\cite{Szabo1996} Thus, truncated methods fail to describe bond dissociation and capture static correlation for molecular systems that approach a distance at which they become non-interacting.

Since quantum computers are poised to solve classically hard problems like the electronic structure of large, strongly correlated molecules and materials, these devices must preserve size consistency under noise. However, noise on current hardware grows with circuit depth and number of qubits, causing the overall error to increase with system size.\cite{Preskill2018,Franca2021,Bharti2022} As a result, the impact of accumulated noise on the preservation of size consistency has not been systematically characterized on quantum hardware.


In the case of non-interacting systems, size consistency can be evaluated on quantum hardware by examining the properties of  additive and multiplicative separability for compound systems composed of  increasing numbers of non-interacting subsystems. Specifically, we consider compound systems composed of non-interacting \ch{H2} molecules. Additive separability is satisfied when the energy per subsystem obtained from the compound system remains constant. Likewise, multiplicative separability requires that the probability of each Slater determinant per subsystem remains constant as additional non-interacting molecules are simulated.

To isolate the effect of device noise on size consistency, we prepare the ground state of the compound system on the device using unitary circuits that preserve these properties in the absence of noise. Here, full configuration interaction (FCI) wavefunctions are used as the exact ground state in the given basis. These FCI states are prepared by a unitary transformation that takes the initial mean-field Hartree-Fock reference state directly into the correlated FCI ground state. Since the FCI wavefunction of a non-interacting system is multiplicatively separable, the corresponding unitary circuits inherit this structure, yielding a size-consistent approach in the absence of noise. Accordingly, the circuit for each subsystem is constructed to be optimally shallow and then combined together to form the circuit of the full compound system.\cite{Khaneja2005,Ashhab2022} Further details are provided in Section 1.1 of the Supporting Information.

The Hartree-Fock reference state of \ch{H2} is computed classically in the Slater-type orbital (STO-3G) basis set using PySCF.\cite{Hehre1969,PySCF2018,PySCF2020} The full electronic Hamiltonian for a single \ch{H2} molecule is mapped onto the device using the Jordan-Wigner transformation on four qubits, one for each spin-orbital. Additionally, by utilizing the $D_{\infty h}$ point group symmetry together with electron number conservation, the qubit count is further reduced to one- and two-qubit representations. Unitary circuits for the preparation of the FCI state, the circuit is independently optimized for each number of qubits.

Once the unitary circuit for each \ch{H2} transforms the Hartree-Fock reference state into the FCI state on the quantum device, the energy of each molecule is obtained by measuring the full quantum circuit using quantum state tomography.\cite{Vogel1989,Hradil1997,James2001,Paris2004} To decrease the number of measurements, we implement a custom Pauli-measurement strategy with classical post-processing. See Section 1.2 of the Supporting Information for additional details. Since the Hamiltonian of the entire compound system is the  sum of the Hamiltonians for each of the $N$ non-interacting \ch{H2} subsystems, its energy is given by the sum of the subsystem expectation values,

\begin{align}
\langle \hat H_{\mathrm{tot}} \rangle = \sum_{n=1}^N  \langle \hat H^{(n)}_{\mathrm{\ch{H2}}}  \rangle
\end{align}

\noindent To measure the energy for each subsystem, each of these Hamiltonians $\hat H^{(n)}_{\mathrm{\ch{H2}}}$ is embedded into the Hilbert space of the full system by tensoring the original Hamiltonian $\hat H_{\mathrm{\ch{H2}}}$ with the identity matrix acting on all other non-interacting subsystems. In the case of $N=2$, $\hat H_{\mathrm{tot}}$ is expressible in any chosen qubit encoding (single-, two-, or four-qubit mappings) as

\begin{align}
\hat H_{\mathrm{tot}} = \hat H_{\mathrm{\ch{H2}}} \otimes \mathds{1} + \mathds{1} \otimes \hat H_{\mathrm{\ch{H2}}}
\end{align}

\noindent Hence, evaluating the energy for each subsystem individually requires the number of measurements to scale linearly. However, because the subsystems are non-interacting, Pauli strings acting on different subsystems commute and can be measured simultaneously.\cite{Verteletskyi2020,DalFavero2025} Thus, by measuring the $N$-fold tensor-product Pauli operator ($P^{\otimes N}$) for each Pauli string in the decomposition of $\hat H_{\mathrm{\ch{H2}}}$,\cite{Nielsen2000} the energy contribution of each subsystem is evaluated from the same measurement data, maintaining a constant number of measurements.
Similar simultaneous measurement and classical post-processing are used for the two- and four-qubit mappings to simultaneously measure Pauli strings in the Hamiltonian that contain commuting Pauli Z operators (e.g., ZIZI, ZZII). Overall, the quantum tomography approaches employed in this work reduce the amount of readout error and ensure that the number of measurements remain constant with system size. The measurement of each Pauli-string is performed using $10^5$ shots.

 \begin{figure*}[t]
 \centering
 \includegraphics[height=6.05cm]{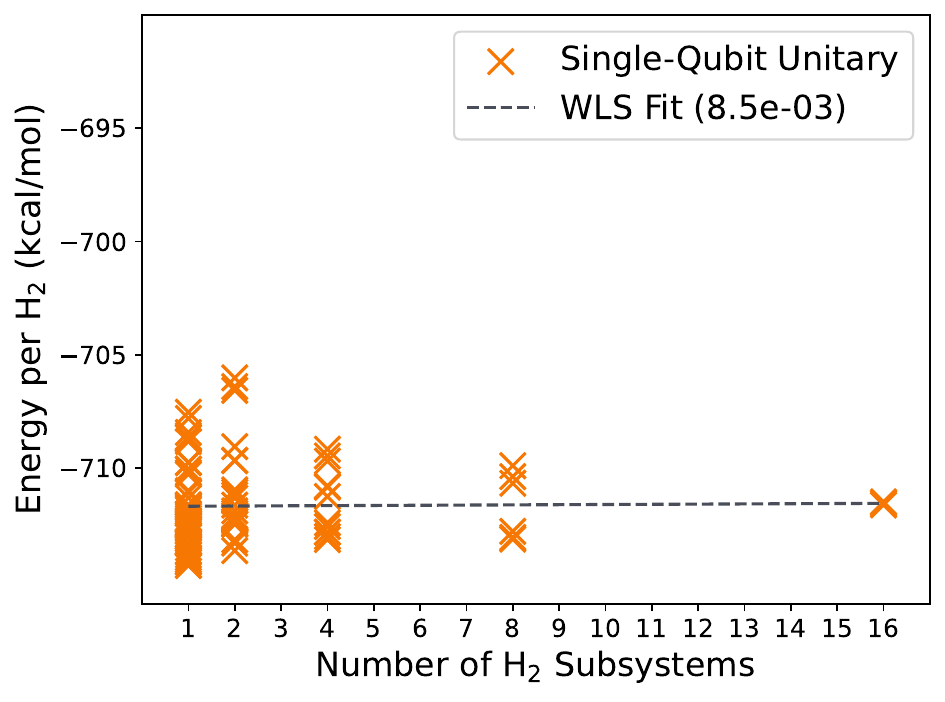}
 \includegraphics[height=6.05cm]{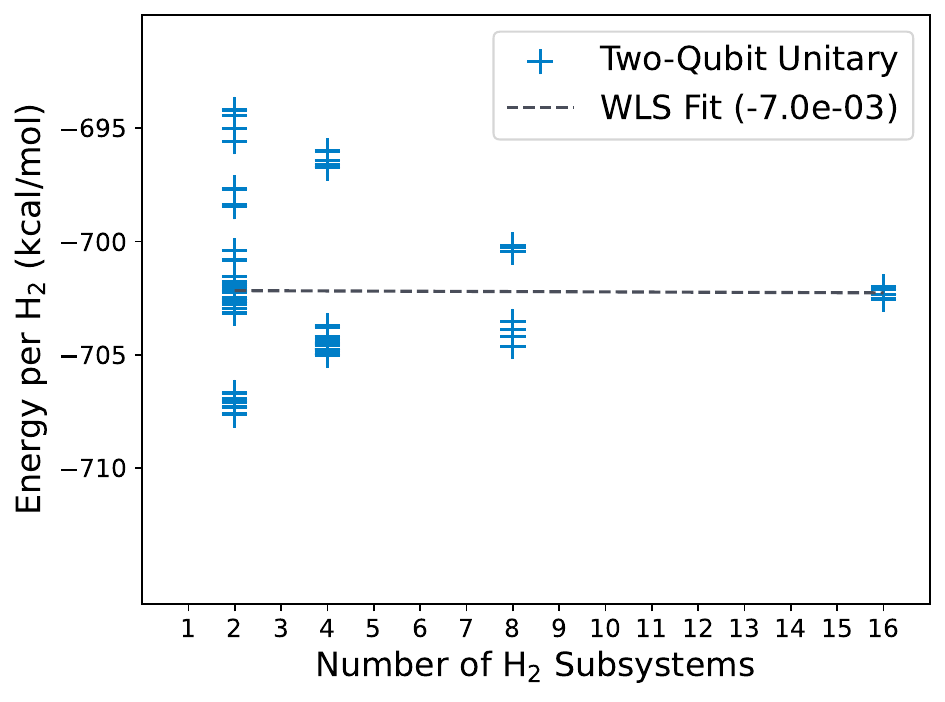}
 \caption{\ Energy per \ch{H2} as system size ($N$) increases. Energies are calculated with single-qubit (left) and two-qubit (right) representations per subsystem. The energies for the single-qubit representation are simulated using the selective sample procedure ($N$=2,4,8,16; $n$=8,4,2,1; $k$=3) with the exception of $N$=1, which is generated using the random sample procedure (s=50). The energies for the two-qubit representation are also modeled using the selective sample procedure ($N$=1,2,4,8; $n$=8,4,2,1; $k$=5). Sample procedures are discussed at the end of the Letter. The energies are reported as the energy per \ch{H2} subsystem, represented as single-qubit: $\times$ (orange) and two-qubit: $+$ (blue) with a weighted least squares (WLS) line: $--$ (gray)}
\label{fgr:1}
\end{figure*}

For each unitary circuit (single-, two-, four-qubit), the maximum number of qubits used to represent the total compound system is sixteen, allowing for the simulation of up to sixteen, eight, and four non-interacting \ch{H2} subsystems for each representation, respectively. For the single- and two-qubit unitary circuits, the average energies per \ch{H2} are calculated to be -711.6 and -702.2 kcal/mol with a standard deviation of 1.82 and 3.48 kcal/mol. The distribution of energies for both unitary circuits, shown in Fig. \ref{fgr:1}, appears to be heteroscedastic and cone-shaped, which arises from the different gate and readout errors for each qubit. Since each qubit has varying degrees of error and susceptibility to noise, the energies from systems with a smaller number of \ch{H2} molecules have a higher variance than those from larger systems, which require at most sixteen qubits. Specifically, larger systems inherently average the difference in error from each qubit during quantum tomography, while smaller systems produce various energies based on the errors of each individual or multiple qubits. This trend can be observed for both single- and two-qubit calculations where the standard deviation of energies begins to decrease as the system size increases.

\begin{table}[t]
\small
  \caption{\ Assessment of Size Consistency}
  \label{tbl:1}
  \begin{tabular*}{0.48\textwidth}{@{\extracolsep{\fill}}crrr}
    \hline
    Qubits per \ch{H2} & \multicolumn{1}{c}{$\Delta$} & \multicolumn{1}{c}{$N_{\mathrm{qubit}}$} & \multicolumn{1}{c}{$N_{\mathrm{\ch{H2}}}$} \\
    \hline
    1 &  8.474 $\times$ 10$^{-3}$ & 118 & 118 \\
    2 & -7.000 $\times$ 10$^{-3}$ & 142 & 71  \\
    4 &  1.221 $\times$ 10$^{-1}$ & 8   & 2   \\
    \hline
  \end{tabular*}

\ Estimated number of qubits ($N_{\mathrm{qubit}}$) and \ch{H2} molecules ($N_{\mathrm{\ch{H2}}}$) within chemical accuracy (1 kcal/mol) as a function of size consistency error ($\Delta$) for single-, two-, and four-qubit unitary circuits. $\Delta$ is measured in units of 1 kcal/mol per qubit.

\end{table}

In order to model heteroscedastic data, a weighted least-squares regression (WLS) line is used to  calculate the slope of energies with different standard deviations. Since the slope is expected to be zero for size-consistent simulations, the WLS line is treated as the error in size consistency ($\Delta$) per \ch{H2}. The $\Delta$ is identified as $8.474 \times 10^{-3}$ kcal/mol per qubit for the single-qubit unitary and $-7.000 \times 10^{-3}$ kcal/mol per qubit for the two-qubit unitary as shown in Table \ref{tbl:1}. Using these slopes, we can estimate that the quantum device preserves size consistency within chemical accuracy for up to 118 and 71 \ch{H2} subsystems (118 and 142 qubits) for the single- and two-qubit representations, respectively. Since the device used in the work, IBM Q Fez, consists of 156 qubits, these results suggest that near-term quantum computers can potentially uphold size consistency across the entire device.

Although a true indication of size consistency would be a slope of zero ($\Delta= 0$), the noise on a quantum computer is not deterministic and may affect the same qubit differently for separate calculations. Additionally, only limited precision can be achieved in a finite number of shots, which will result in a non-zero variance even on noiseless devices. The four-qubit unitary circuit resulted in an average energy per \ch{H2} of -668.3 kcal/mol with a standard deviation of 7.09 kcal/mol (Fig. S1). In this case, the quantum device is estimated to only maintain size consistency within chemical accuracy for up to two \ch{H2} subsystems on a total of eight qubits, with an error of $\Delta = 1.221 \times 10^{-1}$ kcal/mol per qubit. This behavior is consistent with the high multi-qubit error rates on current devices, which increase the variance of measured energies for the four-qubit representation, that has a deeper circuit, relative to those obtained with the single- or two-qubit representation. Since size consistency in this work is assessed by deviations in average energies between compound systems of different sizes, these results suggest that the large $\Delta$ may reflect an insufficient number of samples ($k$) to accurately estimate the true average energy for each compound system under noise. As the circuit depth increases, the noise increases the variance of the measured energies, requiring substantially more samples to accurately estimate the mean and thus size consistency.

\begin{figure*}[h]
 \centering
 \includegraphics[height=6.05cm]{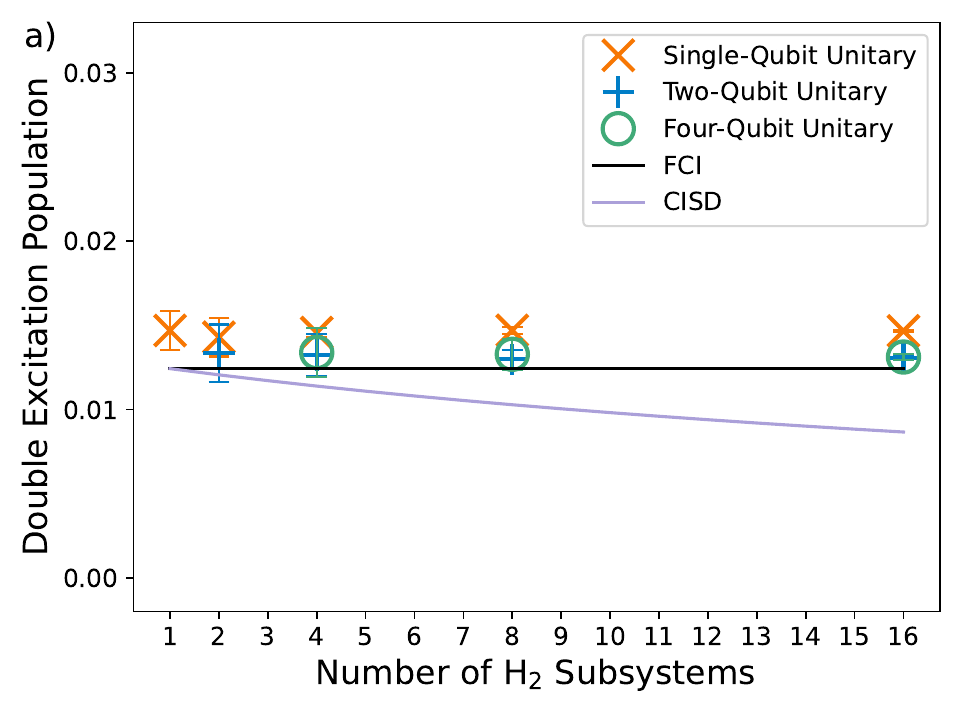}
 \includegraphics[height=6.05cm]{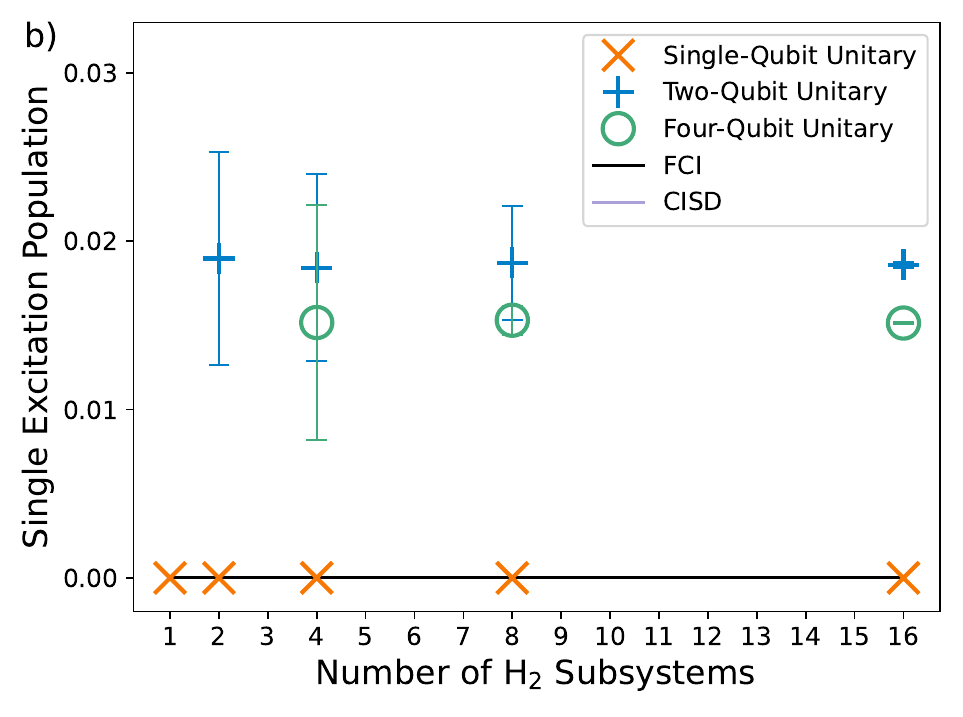}
  \caption{\ \textbf{(a)} Average double-excitation population per \ch{H2} subsystem as system size increases. \textbf{(b)} Average single-excitation population per \ch{H2} as system size increases. The CISD result overlaps exactly with the FCI line. The populations for both figures are represented as single-qubit: $\times$ (orange), two-qubit: $+$ (blue), four-qubit: $\circ$ (green), CISD: $-$ (purple), and FCI: $-$ (black).}
 \label{fgr:2}
\end{figure*}

The average population of the double-excitation from the Hartree-Fock reference for each \ch{H2} is calculated using the probability of measuring each configuration during tomography. The double-excitation population for each unitary circuit is shown in Fig. \ref{fgr:2}a. The results are consistent with the multiplicative separability of the wavefunction since the double-excitation populations are independent of system size. In contrast, CISD produces a curve that underestimates the double-excitation population as system size increases. Additionally, the two- and four-qubit unitary circuits produce populations closer to the FCI populations than the single-qubit unitary in the presence of noise. This suggests that noise predominately affects the probability of measuring the Hartree-Fock reference state because of its larger initial probability. This also demonstrates that as the noise increases for larger systems, the state's correlation can scale properly with the system size.

However, the two- and four-qubit unitary circuits produce less accurate single-excitation populations shown in Fig \ref{fgr:2}b. These populations are solely produced by device noise as the FCI single-excitation population of \ch{H2} in a minimal basis set is zero. CISD also accurately calculates a population of zero as system size increases. In addition, the single-qubit unitary results in single-excitation populations of zero since the single-qubit mapping of the Hamiltonian encodes $|0\rangle$ and $|1\rangle$ as the Hartree-Fock reference state and the double-excitation state, respectively, thus avoiding the appearance of a single-excitation state even in the presence of noise.

Despite the fact that two- and four-qubit circuits produce inaccurate single-excitation populations, these errors do not necessarily prevent quantum computers from performing size-consistent molecular simulations. Additionally, the four-qubit unitary more accurately predicts the probability of single excitation than the two-qubit unitary. This kind of behavior is also seen in Fig. \ref{fgr:2}a, where larger unitary circuits, although more susceptible to noise, produce more accurate populations for higher-order excitations. This can be explained by the large population of number-violating Slater determinants resulting from noise, which are not produced on either single- or two-qubit circuits. These determinants make up a larger population than either population of single or double excitations, which contributes to the large error in energy per \ch{H2} for the four-qubit unitary. Additionally, the same trend observed in Fig. \ref{fgr:1} is demonstrated for single-excitation populations where the standard deviation decreases as the system size increases.

\begin{figure}
 \centering
 \includegraphics[height=6.05cm]{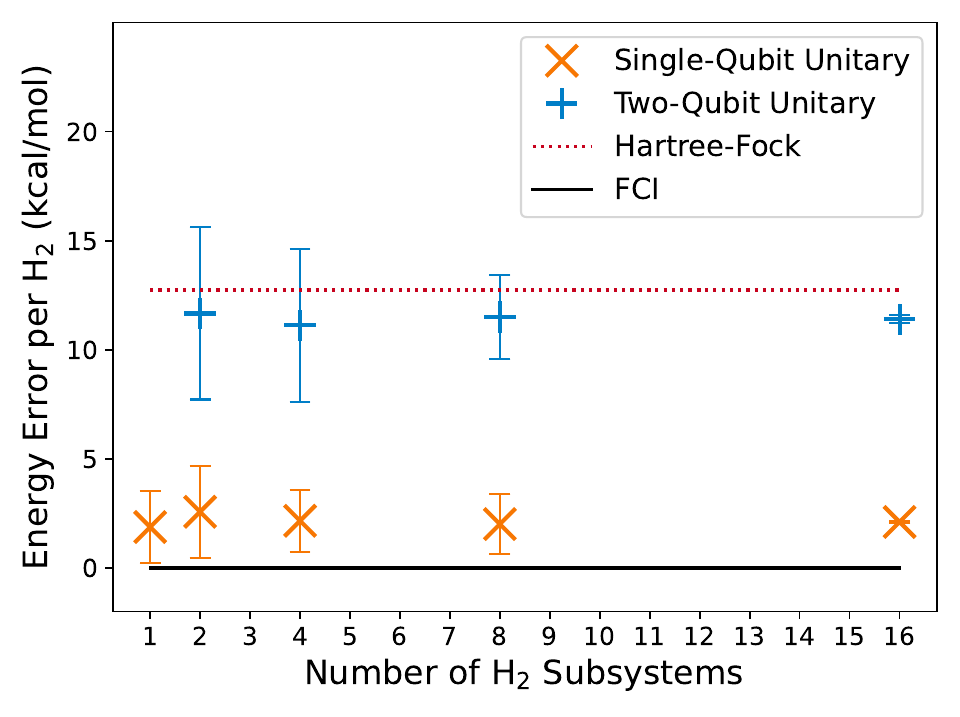}
 \caption{\ Average energy error per \ch{H2} subsystem as system size increases. The energy errors are represented as single-qubit: $\times$ (orange), two-qubit: $+$ (blue), Hartree-Fock: $\cdots$ (red), and FCI: $-$ (black).}
 \label{fgr:3}
\end{figure}

The average error per \ch{H2} for each set of energies is shown in Fig. \ref{fgr:3}. For single- and two-qubit unitary circuits, the average energy error per \ch{H2} is 2.1 and 11.4 kcal/mol. The single-qubit unitary produces a near FCI energy, while the two-qubit unitary is more susceptible to noise, resulting in an energy comparable to Hartree-Fock. This error in energy is due to the incorrect population of single excitations stemming from noise on the device, which leads to a lower population of the Hartree-Fock reference state and subsequently to a higher energy. Additionally, the average energy per \ch{H2} for the four-qubit unitary circuit is 45.2 kcal/mol. While the population of the single-excitation determinant contributes to the error, the dominant source arises due to the measurement of electron number-violating Slater determinants, which suppress the population of the Hartree-Fock reference state. Even though the four-qubit unitary operations produce the FCI state on the device in the absence of noise, readout and gate error shift its energy above the initial Hartree-Fock reference state.


The method-agnostic framework presented in this study demonstrates that current quantum hardware can preserve both additive separability of the energy and multiplicative separability of ground-state excitation populations for non-interacting subsystems. As noise increases for larger subsystems, the device maintains the proper scaling of higher-order excitation populations. In addition, size consistency is observed with high precision for small subsystems, whereas larger subsystems require sufficient measurement statistics to mitigate noise-induced variance. Overall, these results establish that quantum computers have the potential to preserve size consistency within chemical accuracy, even in the presence of hardware noise.

Since size consistency is a fundamental requirement for accurate quantum chemical simulations, these findings provide a necessary foundation for achieving large molecular simulations on quantum computers. Moreover, by directly studying the effect of device noise on size consistency, the observed scaling behavior should extend to a broad class of superconducting and likely other quantum hardware platforms. While further studies are required to establish size extensivity for interacting systems, this work represents a critical first step toward size-consistent quantum algorithms capable of achieving a genuine quantum advantage for modeling strongly correlated systems and periodic structures.

\section{Methods}

By simulating systems of different sizes, each calculation is performed on a different number of qubits, corresponding to various gate and readout errors. To minimize noise and keep error consistent, physical qubits are sorted by CZ gate and readout error from IBM Q Fez calibration data to find the best qubits least susceptible to noise. A custom sampling method is utilized to reproduce the full noise distribution for reduced-qubit systems, enabling a consistent comparison of the energy per subsystem as the system size increases.

Although random selection from a sample set of physical qubits, previously determined by the lowest error, enables the calculation of systems with a non-uniform amount of subsystems (e.g. three or five), producing an accurate distribution of energies requires a large amount of repetitions associated with a high quantum processing runtime cost. Instead, a selective sampling procedure is used to select different qubits for each calculation, which are sampled from the entire set of qubits. For example, consider a compound system composed of sixteen single-qubit subsystems simulated on sixteen qubits. In order to reproduce the individual gate and readout error of this full system for smaller systems (four or eight subsystems), each calculation is repeated and selectively sampled over the set of all physical qubits (e.g. two single-qubit subsystems simulated on two qubits, sampled 8 times).

Additionally, for greater accuracy, the entire sample of calculations simulated on each of the sixteen qubits is treated as a set, which is repeated $k$ times, forming a total sample population spanning $n$ samples and $k$ sets. This procedure reduces the number of samples necessary to produce an accurate distribution, performing nearly identically to the random selection procedure in the high sampling limit. In general, this procedure measures the energy per subsystem on each qubit $k$ times for each compound system composed of $N$ subsystems, where $n$ is the number of samples required to measure each system on each qubit ($n=16/N$).

\begin{acknowledgement}

D.A.M gratefully acknowledges support from the U.S. National Science Foundation Grant No. CHE-2155082 as well as support from IBM under the IBM-UChicago Quantum Collaboration including IBM Quantum services. The views expressed are those of the authors, and do not reflect the official policy or position of IBM or the IBM Quantum team.


\end{acknowledgement}

\section{Data Availability}
The python code is publicly available on GitHub: \href{https://github.com/damazz/Size-Consistent-Quantum-Chemistry}{https://github.com/damazz/Size-Consistent-Quantum-Chemistry}. The calibration data for IBM Q Fez is available upon request.

\begin{suppinfo}

The Supporting Information is available free of charge.

\end{suppinfo}

\bibliography{achemso-demo}

\providecommand{\latin}[1]{#1}
\makeatletter
\providecommand{\doi}
  {\begingroup\let\do\@makeother\dospecials
  \catcode`\{=1 \catcode`\}=2 \doi@aux}
\providecommand{\doi@aux}[1]{\endgroup\texttt{#1}}
\makeatother
\providecommand*\mcitethebibliography{\thebibliography}
\csname @ifundefined\endcsname{endmcitethebibliography}
  {\let\endmcitethebibliography\endthebibliography}{}
\begin{mcitethebibliography}{57}
\providecommand*\natexlab[1]{#1}
\providecommand*\mciteSetBstSublistMode[1]{}
\providecommand*\mciteSetBstMaxWidthForm[2]{}
\providecommand*\mciteBstWouldAddEndPuncttrue
  {\def\EndOfBibitem{\unskip.}}
\providecommand*\mciteBstWouldAddEndPunctfalse
  {\let\EndOfBibitem\relax}
\providecommand*\mciteSetBstMidEndSepPunct[3]{}
\providecommand*\mciteSetBstSublistLabelBeginEnd[3]{}
\providecommand*\EndOfBibitem{}
\mciteSetBstSublistMode{f}
\mciteSetBstMaxWidthForm{subitem}{(\alph{mcitesubitemcount})}
\mciteSetBstSublistLabelBeginEnd
  {\mcitemaxwidthsubitemform\space}
  {\relax}
  {\relax}

\bibitem[Pople \latin{et~al.}(1976)Pople, Binkley, and Seeger]{Pople1976}
Pople,~J.~A.; Binkley,~J.~S.; Seeger,~R. Theoretical models incorporating
  electron correlation. \emph{Int. J. Quantum Chem.} \textbf{1976}, \emph{10},
  1--19\relax
\mciteBstWouldAddEndPuncttrue
\mciteSetBstMidEndSepPunct{\mcitedefaultmidpunct}
{\mcitedefaultendpunct}{\mcitedefaultseppunct}\relax
\EndOfBibitem
\bibitem[Bartlett and Shavitt(1977)Bartlett, and Shavitt]{Bartlett1977}
Bartlett,~R.~J.; Shavitt,~I. Determination of the size-consistency error in the
  single and double excitation configuration interaction model. \emph{Int. J.
  Quantum Chem.} \textbf{1977}, \emph{12}, 165--173\relax
\mciteBstWouldAddEndPuncttrue
\mciteSetBstMidEndSepPunct{\mcitedefaultmidpunct}
{\mcitedefaultendpunct}{\mcitedefaultseppunct}\relax
\EndOfBibitem
\bibitem[Bartlett(1981)]{Barlett1981}
Bartlett,~R.~J. Many-Body Perturbation Theory and Coupled Cluster Theory for
  Electron Correlation in Molecules. \emph{Annu. Rev. Phys. Chem.}
  \textbf{1981}, \emph{32}, 359--401\relax
\mciteBstWouldAddEndPuncttrue
\mciteSetBstMidEndSepPunct{\mcitedefaultmidpunct}
{\mcitedefaultendpunct}{\mcitedefaultseppunct}\relax
\EndOfBibitem
\bibitem[Cremer and Thiel(1987)Cremer, and Thiel]{Cremer1987}
Cremer,~D.; Thiel,~W. On the importance of size-consistency corrections in
  semiempirical MNDOC calculations. \emph{J. Comput. Chem.} \textbf{1987},
  \emph{8}, 48--50\relax
\mciteBstWouldAddEndPuncttrue
\mciteSetBstMidEndSepPunct{\mcitedefaultmidpunct}
{\mcitedefaultendpunct}{\mcitedefaultseppunct}\relax
\EndOfBibitem
\bibitem[Deleuze and Pickup(1995)Deleuze, and Pickup]{Deleuze1995}
Deleuze,~M.; Pickup,~B.~T. Size consistency and size extensivity of
  linear-response properties using the perturbed electron propagator. \emph{J.
  Chem. Phys.} \textbf{1995}, \emph{102}, 8967--8977\relax
\mciteBstWouldAddEndPuncttrue
\mciteSetBstMidEndSepPunct{\mcitedefaultmidpunct}
{\mcitedefaultendpunct}{\mcitedefaultseppunct}\relax
\EndOfBibitem
\bibitem[Sheppard(1984)]{Sheppard1984}
Sheppard,~M.~G. Size consistency in perturbation theories. \emph{J. Chem.
  Phys.} \textbf{1984}, \emph{80}, 1225--1229\relax
\mciteBstWouldAddEndPuncttrue
\mciteSetBstMidEndSepPunct{\mcitedefaultmidpunct}
{\mcitedefaultendpunct}{\mcitedefaultseppunct}\relax
\EndOfBibitem
\bibitem[Szabo and Ostlund(1996)Szabo, and Ostlund]{Szabo1996}
Szabo,~A.; Ostlund,~N.~S. \emph{Modern Quantum Chemistry: Introduction to
  Advanced Electronic Structure Theory}, 1st ed.; Dover Publications, Inc.:
  Mineola, 1996\relax
\mciteBstWouldAddEndPuncttrue
\mciteSetBstMidEndSepPunct{\mcitedefaultmidpunct}
{\mcitedefaultendpunct}{\mcitedefaultseppunct}\relax
\EndOfBibitem
\bibitem[Mahapatra \latin{et~al.}(1998)Mahapatra, Datta, and
  Mukherjee]{Mahapatra1998}
Mahapatra,~U.~S.; Datta,~B.; Mukherjee,~D. A state-specific multi-reference
  coupled cluster formalism with molecular applications. \emph{Mol. Phys.}
  \textbf{1998}, \emph{94}, 157--171\relax
\mciteBstWouldAddEndPuncttrue
\mciteSetBstMidEndSepPunct{\mcitedefaultmidpunct}
{\mcitedefaultendpunct}{\mcitedefaultseppunct}\relax
\EndOfBibitem
\bibitem[Mahapatra \latin{et~al.}(1999)Mahapatra, Datta, and
  Mukherjee]{Mahapatra1999}
Mahapatra,~U.~S.; Datta,~B.; Mukherjee,~D. A size-consistent state-specific
  multireference coupled cluster theory: Formal developments and molecular
  applications. \emph{J. Chem. Phys.} \textbf{1999}, \emph{110},
  6171--6188\relax
\mciteBstWouldAddEndPuncttrue
\mciteSetBstMidEndSepPunct{\mcitedefaultmidpunct}
{\mcitedefaultendpunct}{\mcitedefaultseppunct}\relax
\EndOfBibitem
\bibitem[Deleuze(2003)]{Deleuze2003}
Deleuze,~M.~S. The issues of size and charge consistency and the implications
  of translation symmetry in advanced Green's function theories. \emph{Int. J.
  Quantum Chem.} \textbf{2003}, \emph{93}, 191--211\relax
\mciteBstWouldAddEndPuncttrue
\mciteSetBstMidEndSepPunct{\mcitedefaultmidpunct}
{\mcitedefaultendpunct}{\mcitedefaultseppunct}\relax
\EndOfBibitem
\bibitem[Nooijen \latin{et~al.}(2005)Nooijen, Shamasundar, and
  Mukherjee]{Nooijen2005}
Nooijen,~M.; Shamasundar,~K.~R.; Mukherjee,~D. Reflections on size-extensivity,
  size-consistency and generalized extensivity in many-body theory. \emph{Mol.
  Phys.} \textbf{2005}, \emph{103}, 2277--2298\relax
\mciteBstWouldAddEndPuncttrue
\mciteSetBstMidEndSepPunct{\mcitedefaultmidpunct}
{\mcitedefaultendpunct}{\mcitedefaultseppunct}\relax
\EndOfBibitem
\bibitem[Hirata(2011)]{Hirata2011}
Hirata,~S. Thermodynamic limit and size-consistent design. \emph{Theor. Chem.
  Acc.} \textbf{2011}, \emph{129}, 727--746\relax
\mciteBstWouldAddEndPuncttrue
\mciteSetBstMidEndSepPunct{\mcitedefaultmidpunct}
{\mcitedefaultendpunct}{\mcitedefaultseppunct}\relax
\EndOfBibitem
\bibitem[Sen \latin{et~al.}(2015)Sen, Sen, Samanta, and Mukherjee]{Sen2015}
Sen,~A.; Sen,~S.; Samanta,~P.~K.; Mukherjee,~D. Unitary group adapted state
  specific multireference perturbation theory: Formulation and pilot
  applications. \emph{J. Comput. Chem.} \textbf{2015}, \emph{36},
  670--688\relax
\mciteBstWouldAddEndPuncttrue
\mciteSetBstMidEndSepPunct{\mcitedefaultmidpunct}
{\mcitedefaultendpunct}{\mcitedefaultseppunct}\relax
\EndOfBibitem
\bibitem[Mih\'alka \latin{et~al.}(2020)Mih\'alka, Surj\'an, and
  Szabados]{Mihalka2020}
Mih\'alka,~Z.~E.; Surj\'an,~P.~R.; Szabados,~A. Half-Projection of the Strongly
  Orthogonal Unrestricted Geminals’ Product Wave Function. \emph{J. Chem.
  Theory Comput.} \textbf{2020}, \emph{16}, 892--903\relax
\mciteBstWouldAddEndPuncttrue
\mciteSetBstMidEndSepPunct{\mcitedefaultmidpunct}
{\mcitedefaultendpunct}{\mcitedefaultseppunct}\relax
\EndOfBibitem
\bibitem[Xiao \latin{et~al.}(2024)Xiao, Zhao, Ren, Fang, and Li]{Xiao2024}
Xiao,~X.; Zhao,~H.; Ren,~J.; Fang,~W.-H.; Li,~Z. Physics-Constrained
  Hardware-Efficient Ansatz on Quantum Computers That Is Universal,
  Systematically Improvable, and Size-Consistent. \emph{J. Chem. Theory
  Comput.} \textbf{2024}, \emph{20}, 1912--1922\relax
\mciteBstWouldAddEndPuncttrue
\mciteSetBstMidEndSepPunct{\mcitedefaultmidpunct}
{\mcitedefaultendpunct}{\mcitedefaultseppunct}\relax
\EndOfBibitem
\bibitem[Kong(2024)]{Kong2024}
Kong,~J. Density functional theory for fractional charge: Locality, size
  consistency, and exchange-correlation. \emph{J. Chem. Phys.} \textbf{2024},
  \emph{161}, 224111\relax
\mciteBstWouldAddEndPuncttrue
\mciteSetBstMidEndSepPunct{\mcitedefaultmidpunct}
{\mcitedefaultendpunct}{\mcitedefaultseppunct}\relax
\EndOfBibitem
\bibitem[Bardeen \latin{et~al.}(1957)Bardeen, Cooper, and
  Schrieffer]{Bardeen1957}
Bardeen,~J.; Cooper,~L.~N.; Schrieffer,~J.~R. Theory of Superconductivity.
  \emph{Phys. Rev.} \textbf{1957}, \emph{108}, 1175--1204\relax
\mciteBstWouldAddEndPuncttrue
\mciteSetBstMidEndSepPunct{\mcitedefaultmidpunct}
{\mcitedefaultendpunct}{\mcitedefaultseppunct}\relax
\EndOfBibitem
\bibitem[Dagotto(1994)]{Dagotto1994}
Dagotto,~E. Correlated electrons in high-temperature superconductors.
  \emph{Rev. Mod. Phys.} \textbf{1994}, \emph{66}, 763--840\relax
\mciteBstWouldAddEndPuncttrue
\mciteSetBstMidEndSepPunct{\mcitedefaultmidpunct}
{\mcitedefaultendpunct}{\mcitedefaultseppunct}\relax
\EndOfBibitem
\bibitem[Bennemann and Ketterson(2004)Bennemann, and Ketterson]{Bennemann2004}
Bennemann,~K.; Ketterson,~J. \emph{The Physics of Superconductors: Vol. II.
  Superconductivity in Nanostructures, High-Tc and Novel Superconductors,
  Organic Superconductors}; Springer, 2004\relax
\mciteBstWouldAddEndPuncttrue
\mciteSetBstMidEndSepPunct{\mcitedefaultmidpunct}
{\mcitedefaultendpunct}{\mcitedefaultseppunct}\relax
\EndOfBibitem
\bibitem[Lee \latin{et~al.}(2006)Lee, Nagaosa, and Wen]{Lee2006}
Lee,~P.; Nagaosa,~N.; Wen,~X. Doping a Mott insulator: Physics of
  high-temperature superconductivity. \emph{Rev. Mod. Phys.} \textbf{2006},
  \emph{78}, 17--85\relax
\mciteBstWouldAddEndPuncttrue
\mciteSetBstMidEndSepPunct{\mcitedefaultmidpunct}
{\mcitedefaultendpunct}{\mcitedefaultseppunct}\relax
\EndOfBibitem
\bibitem[Giovannetti \latin{et~al.}(2004)Giovannetti, Lloyd, and
  Maccone]{Giovannetti2004}
Giovannetti,~V.; Lloyd,~S.; Maccone,~L. Quantum-enhanced measurements: Beating
  the standard quantum limit. \emph{Science} \textbf{2004}, \emph{306},
  1330--1336\relax
\mciteBstWouldAddEndPuncttrue
\mciteSetBstMidEndSepPunct{\mcitedefaultmidpunct}
{\mcitedefaultendpunct}{\mcitedefaultseppunct}\relax
\EndOfBibitem
\bibitem[Taylor \latin{et~al.}(2008)Taylor, Cappellaro, Childress, Jiang,
  Budker, Hemmer, Yacoby, Walsworth, and Lukin]{Talor2008}
Taylor,~J.~M.; Cappellaro,~P.; Childress,~L.; Jiang,~L.; Budker,~D.;
  Hemmer,~P.~R.; Yacoby,~A.; Walsworth,~R.; Lukin,~M.~D. High-sensitivity
  diamond magnetometer with nanoscale resolution. \emph{Nat. Phys.}
  \textbf{2008}, \emph{4}, 810--816\relax
\mciteBstWouldAddEndPuncttrue
\mciteSetBstMidEndSepPunct{\mcitedefaultmidpunct}
{\mcitedefaultendpunct}{\mcitedefaultseppunct}\relax
\EndOfBibitem
\bibitem[Degen \latin{et~al.}(2017)Degen, Reinhard, and Cappellaro]{Degen2017}
Degen,~C.~L.; Reinhard,~F.; Cappellaro,~P. Quantum sensing. \emph{Rev. Mod.
  Phys.} \textbf{2017}, \emph{89}, 035002\relax
\mciteBstWouldAddEndPuncttrue
\mciteSetBstMidEndSepPunct{\mcitedefaultmidpunct}
{\mcitedefaultendpunct}{\mcitedefaultseppunct}\relax
\EndOfBibitem
\bibitem[Ostroverkhova(2016)]{Ostroverkhova2016}
Ostroverkhova,~O. Organic Optoelectronic Materials: Mechanisms and
  Applications. \emph{Chem. Rev.} \textbf{2016}, \emph{116}, 13279--13412\relax
\mciteBstWouldAddEndPuncttrue
\mciteSetBstMidEndSepPunct{\mcitedefaultmidpunct}
{\mcitedefaultendpunct}{\mcitedefaultseppunct}\relax
\EndOfBibitem
\bibitem[Dagotto(2005)]{Dagotto2005}
Dagotto,~E. Complexity in strongly correlated electronic systems.
  \emph{Science} \textbf{2005}, \emph{309}, 257--262\relax
\mciteBstWouldAddEndPuncttrue
\mciteSetBstMidEndSepPunct{\mcitedefaultmidpunct}
{\mcitedefaultendpunct}{\mcitedefaultseppunct}\relax
\EndOfBibitem
\bibitem[Mazziotti(2007)]{Mazziotti2007.3}
Mazziotti,~D.~A. \emph{Reduced‐Density‐Matrix Mechanics: With Application
  to Many‐Electron Atoms and Molecules}; John Wiley \& Sons, Ltd, 2007;
  Chapter 8, pp 165--203\relax
\mciteBstWouldAddEndPuncttrue
\mciteSetBstMidEndSepPunct{\mcitedefaultmidpunct}
{\mcitedefaultendpunct}{\mcitedefaultseppunct}\relax
\EndOfBibitem
\bibitem[Ganoe and Shee(2024)Ganoe, and Shee]{Ganoe2024}
Ganoe,~B.; Shee,~J. On the notion of strong correlation in electronic structure
  theory. \emph{Faraday Discuss.} \textbf{2024}, \emph{254}, 53--75\relax
\mciteBstWouldAddEndPuncttrue
\mciteSetBstMidEndSepPunct{\mcitedefaultmidpunct}
{\mcitedefaultendpunct}{\mcitedefaultseppunct}\relax
\EndOfBibitem
\bibitem[Schouten \latin{et~al.}(2025)Schouten, Ewing, and
  Mazziotti]{Schouten2025}
Schouten,~A.~O.; Ewing,~S.; Mazziotti,~D.~A. Two-electron reduced density
  matrix method for the electronic structure of correlated materials.
  \emph{Phys. Rev. B} \textbf{2025}, \emph{112}, 235130\relax
\mciteBstWouldAddEndPuncttrue
\mciteSetBstMidEndSepPunct{\mcitedefaultmidpunct}
{\mcitedefaultendpunct}{\mcitedefaultseppunct}\relax
\EndOfBibitem
\bibitem[Roos(1980)]{Roos1980}
Roos,~B.~O. The complete active space SCF method in a fock-matrix-based
  super-CI formulation. \emph{Int. J. Quantum Chem.} \textbf{1980}, \emph{18},
  175--189\relax
\mciteBstWouldAddEndPuncttrue
\mciteSetBstMidEndSepPunct{\mcitedefaultmidpunct}
{\mcitedefaultendpunct}{\mcitedefaultseppunct}\relax
\EndOfBibitem
\bibitem[Roos(1987)]{Roos1987}
Roos,~B.~O. \emph{Adv. Chem. Phys.}; John Wiley \& Sons, Ltd, 1987; Chapter 7,
  pp 399--445\relax
\mciteBstWouldAddEndPuncttrue
\mciteSetBstMidEndSepPunct{\mcitedefaultmidpunct}
{\mcitedefaultendpunct}{\mcitedefaultseppunct}\relax
\EndOfBibitem
\bibitem[Feynman(1982)]{Feynman1982}
Feynman,~R.~P. Simulating physics with computers. \emph{Int. J. Theor. Phys.}
  \textbf{1982}, \emph{21}, 467--488\relax
\mciteBstWouldAddEndPuncttrue
\mciteSetBstMidEndSepPunct{\mcitedefaultmidpunct}
{\mcitedefaultendpunct}{\mcitedefaultseppunct}\relax
\EndOfBibitem
\bibitem[McArdle \latin{et~al.}(2020)McArdle, Endo, Aspuru-Guzik, Benjamin, and
  Yuan]{McArdle2020}
McArdle,~S.; Endo,~S.; Aspuru-Guzik,~A.; Benjamin,~S.~C.; Yuan,~X. Quantum
  computational chemistry. \emph{Rev. Mod. Phys.} \textbf{2020}, \emph{92},
  015003\relax
\mciteBstWouldAddEndPuncttrue
\mciteSetBstMidEndSepPunct{\mcitedefaultmidpunct}
{\mcitedefaultendpunct}{\mcitedefaultseppunct}\relax
\EndOfBibitem
\bibitem[Motta and Rice(2022)Motta, and Rice]{Motta2022}
Motta,~M.; Rice,~J.~E. Emerging quantum computing algorithms for quantum
  chemistry. \emph{WIREs Comput. Mol. Sci.} \textbf{2022}, \emph{12},
  e1580\relax
\mciteBstWouldAddEndPuncttrue
\mciteSetBstMidEndSepPunct{\mcitedefaultmidpunct}
{\mcitedefaultendpunct}{\mcitedefaultseppunct}\relax
\EndOfBibitem
\bibitem[Smart and Mazziotti(2022)Smart, and Mazziotti]{Smart2022.6}
Smart,~S.~E.; Mazziotti,~D.~A. Many-fermion simulation from the contracted
  quantum eigensolver without fermionic encoding of the wave function.
  \emph{Phys. Rev. A} \textbf{2022}, \emph{105}, 062424\relax
\mciteBstWouldAddEndPuncttrue
\mciteSetBstMidEndSepPunct{\mcitedefaultmidpunct}
{\mcitedefaultendpunct}{\mcitedefaultseppunct}\relax
\EndOfBibitem
\bibitem[Smart and Mazziotti(2022)Smart, and Mazziotti]{Smart2022.9}
Smart,~S.~E.; Mazziotti,~D.~A. Accelerated Convergence of Contracted Quantum
  Eigensolvers through a Quasi-Second-Order, Locally Parameterized
  Optimization. \emph{J. Chem. Theory Comput.} \textbf{2022}, \emph{18},
  5286--5296\relax
\mciteBstWouldAddEndPuncttrue
\mciteSetBstMidEndSepPunct{\mcitedefaultmidpunct}
{\mcitedefaultendpunct}{\mcitedefaultseppunct}\relax
\EndOfBibitem
\bibitem[Wang and Mazziotti(2023)Wang, and Mazziotti]{Wang2023.9}
Wang,~Y.; Mazziotti,~D.~A. Electronic excited states from a variance-based
  contracted quantum eigensolver. \emph{Phys. Rev. A} \textbf{2023},
  \emph{108}, 022814\relax
\mciteBstWouldAddEndPuncttrue
\mciteSetBstMidEndSepPunct{\mcitedefaultmidpunct}
{\mcitedefaultendpunct}{\mcitedefaultseppunct}\relax
\EndOfBibitem
\bibitem[Wang \latin{et~al.}(2023)Wang, Sager-Smith, and
  Mazziotti]{Wang2023.10}
Wang,~Y.; Sager-Smith,~L.~M.; Mazziotti,~D.~A. Quantum simulation of bosons
  with the contracted quantum eigensolver. \emph{New J. Phys.} \textbf{2023},
  \emph{25}, 103005\relax
\mciteBstWouldAddEndPuncttrue
\mciteSetBstMidEndSepPunct{\mcitedefaultmidpunct}
{\mcitedefaultendpunct}{\mcitedefaultseppunct}\relax
\EndOfBibitem
\bibitem[Cao \latin{et~al.}(2019)Cao, Romero, Olson, Degroote, Johnson,
  Kieferová, Kivlichan, Menke, Peropadre, Sawaya, Sim, Veis, and
  Aspuru-Guzik]{Cao2019}
Cao,~Y.; Romero,~J.; Olson,~J.~P.; Degroote,~M.; Johnson,~P.~D.;
  Kieferová,~M.; Kivlichan,~I.~D.; Menke,~T.; Peropadre,~B.; Sawaya,~N. P.~D.;
  Sim,~S.; Veis,~L.; Aspuru-Guzik,~A. Quantum Chemistry in the Age of Quantum
  Computing. \emph{Chem. Rev.} \textbf{2019}, \emph{119}, 10856--10915\relax
\mciteBstWouldAddEndPuncttrue
\mciteSetBstMidEndSepPunct{\mcitedefaultmidpunct}
{\mcitedefaultendpunct}{\mcitedefaultseppunct}\relax
\EndOfBibitem
\bibitem[Bauer \latin{et~al.}(2020)Bauer, Bravyi, Motta, and Chan]{Bauer2020}
Bauer,~B.; Bravyi,~S.; Motta,~M.; Chan,~G. K.-L. Quantum Algorithms for Quantum
  Chemistry and Quantum Materials Science. \emph{Chem. Rev.} \textbf{2020},
  \emph{120}, 12685--12717\relax
\mciteBstWouldAddEndPuncttrue
\mciteSetBstMidEndSepPunct{\mcitedefaultmidpunct}
{\mcitedefaultendpunct}{\mcitedefaultseppunct}\relax
\EndOfBibitem
\bibitem[Preskill(2018)]{Preskill2018}
Preskill,~J. Quantum Computing in the NISQ era and beyond. \emph{Quantum}
  \textbf{2018}, \emph{2}, 79\relax
\mciteBstWouldAddEndPuncttrue
\mciteSetBstMidEndSepPunct{\mcitedefaultmidpunct}
{\mcitedefaultendpunct}{\mcitedefaultseppunct}\relax
\EndOfBibitem
\bibitem[Makov and Payne(1995)Makov, and Payne]{Makov1995}
Makov,~G.; Payne,~M.~C. Periodic boundary-conditions in ab-initio calculations.
  \emph{Phys. Rev. B} \textbf{1995}, \emph{51}, 4014--4022\relax
\mciteBstWouldAddEndPuncttrue
\mciteSetBstMidEndSepPunct{\mcitedefaultmidpunct}
{\mcitedefaultendpunct}{\mcitedefaultseppunct}\relax
\EndOfBibitem
\bibitem[Kratzer and Neugebauer(2019)Kratzer, and Neugebauer]{Kratzer2019}
Kratzer,~P.; Neugebauer,~J. The Basics of Electronic Structure Theory for
  Periodic Systems. \emph{Front. Chem.} \textbf{2019}, \emph{7}, 106\relax
\mciteBstWouldAddEndPuncttrue
\mciteSetBstMidEndSepPunct{\mcitedefaultmidpunct}
{\mcitedefaultendpunct}{\mcitedefaultseppunct}\relax
\EndOfBibitem
\bibitem[Stilck~Fran\c{c}a and Garc\'{i}a-Patr\'{o}n(2021)Stilck~Fran\c{c}a,
  and Garc\'{i}a-Patr\'{o}n]{Franca2021}
Stilck~Fran\c{c}a,~D.; Garc\'{i}a-Patr\'{o}n,~R. Limitations of optimization
  algorithms on noisy quantum devices. \emph{Nat. Phys.} \textbf{2021},
  \emph{17}, 1221--1227\relax
\mciteBstWouldAddEndPuncttrue
\mciteSetBstMidEndSepPunct{\mcitedefaultmidpunct}
{\mcitedefaultendpunct}{\mcitedefaultseppunct}\relax
\EndOfBibitem
\bibitem[Bharti \latin{et~al.}(2022)Bharti, Cervera-Lierta, Kyaw, Haug,
  Alperin-Lea, Anand, Degroote, Heimonen, Kottmann, Menke, Mok, Sim, Kwek, and
  Aspuru-Guzik]{Bharti2022}
Bharti,~K.; Cervera-Lierta,~A.; Kyaw,~T.~H.; Haug,~T.; Alperin-Lea,~S.;
  Anand,~A.; Degroote,~M.; Heimonen,~H.; Kottmann,~J.~S.; Menke,~T.;
  Mok,~W.-K.; Sim,~S.; Kwek,~L.-C.; Aspuru-Guzik,~A. Noisy intermediate-scale
  quantum algorithms. \emph{Rev. Mod. Phys.} \textbf{2022}, \emph{94},
  015004\relax
\mciteBstWouldAddEndPuncttrue
\mciteSetBstMidEndSepPunct{\mcitedefaultmidpunct}
{\mcitedefaultendpunct}{\mcitedefaultseppunct}\relax
\EndOfBibitem
\bibitem[Khaneja \latin{et~al.}(2005)Khaneja, Reiss, Kehlet,
  Schulte-Herbrüggen, and Glaser]{Khaneja2005}
Khaneja,~N.; Reiss,~T.; Kehlet,~C.; Schulte-Herbrüggen,~T.; Glaser,~S.~J.
  Optimal control of coupled spin dynamics: design of NMR pulse sequences by
  gradient ascent algorithms. \emph{J. Magn. Reson.} \textbf{2005}, \emph{172},
  296--305\relax
\mciteBstWouldAddEndPuncttrue
\mciteSetBstMidEndSepPunct{\mcitedefaultmidpunct}
{\mcitedefaultendpunct}{\mcitedefaultseppunct}\relax
\EndOfBibitem
\bibitem[Ashhab \latin{et~al.}(2022)Ashhab, Yamamoto, Yoshihara, and
  Semba]{Ashhab2022}
Ashhab,~S.; Yamamoto,~N.; Yoshihara,~F.; Semba,~K. Numerical analysis of
  quantum circuits for state preparation and unitary operator synthesis.
  \emph{Phys. Rev. A} \textbf{2022}, \emph{106}, 022426\relax
\mciteBstWouldAddEndPuncttrue
\mciteSetBstMidEndSepPunct{\mcitedefaultmidpunct}
{\mcitedefaultendpunct}{\mcitedefaultseppunct}\relax
\EndOfBibitem
\bibitem[Hehre \latin{et~al.}(1969)Hehre, Stewart, and Pople]{Hehre1969}
Hehre,~W.~J.; Stewart,~R.~F.; Pople,~J.~A. Self‐Consistent
  Molecular‐Orbital Methods. I. Use of Gaussian Expansions of Slater‐Type
  Atomic Orbitals. \emph{J. Chem. Phys.} \textbf{1969}, \emph{51},
  2657--2664\relax
\mciteBstWouldAddEndPuncttrue
\mciteSetBstMidEndSepPunct{\mcitedefaultmidpunct}
{\mcitedefaultendpunct}{\mcitedefaultseppunct}\relax
\EndOfBibitem
\bibitem[Sun \latin{et~al.}(2018)Sun, Berkelbach, Blunt, Booth, Guo, Li, Liu,
  McClain, Sayfutyarova, Sharma, Wouters, and Chan]{PySCF2018}
Sun,~Q.; Berkelbach,~T.~C.; Blunt,~N.~S.; Booth,~G.~H.; Guo,~S.; Li,~Z.;
  Liu,~J.; McClain,~J.~D.; Sayfutyarova,~E.~R.; Sharma,~S.; Wouters,~S.;
  Chan,~G. K.-L. PySCF: the Python-based simulations of chemistry framework.
  \emph{WIREs Comput. Mol. Sci.} \textbf{2018}, \emph{8}, e1340\relax
\mciteBstWouldAddEndPuncttrue
\mciteSetBstMidEndSepPunct{\mcitedefaultmidpunct}
{\mcitedefaultendpunct}{\mcitedefaultseppunct}\relax
\EndOfBibitem
\bibitem[Sun \latin{et~al.}(2020)Sun, Zhang, Banerjee, Bao, Barbry, Blunt,
  Bogdanov, Booth, Chen, Cui, Eriksen, Gao, Guo, Hermann, Hermes, Koh, Koval,
  Lehtola, Li, Liu, Mardirossian, McClain, Motta, Mussard, Pham, Pulkin,
  Purwanto, Robinson, Ronca, Sayfutyarova, Scheurer, Schurkus, Smith, Sun, Sun,
  Upadhyay, Wagner, Wang, White, Whitfield, Williamson, Wouters, Yang, Yu, Zhu,
  Berkelbach, Sharma, Sokolov, and Chan]{PySCF2020}
Sun,~Q. \latin{et~al.}  Recent developments in the PySCF program package.
  \emph{J. Chem. Phys.} \textbf{2020}, \emph{153}, 024109\relax
\mciteBstWouldAddEndPuncttrue
\mciteSetBstMidEndSepPunct{\mcitedefaultmidpunct}
{\mcitedefaultendpunct}{\mcitedefaultseppunct}\relax
\EndOfBibitem
\bibitem[Vogel and Risken(1989)Vogel, and Risken]{Vogel1989}
Vogel,~K.; Risken,~H. Determination of quasiprobability distributions in terms
  of probability distributions for the rotated quadrature phase. \emph{Phys.
  Rev. A} \textbf{1989}, \emph{40}, 2847--2849\relax
\mciteBstWouldAddEndPuncttrue
\mciteSetBstMidEndSepPunct{\mcitedefaultmidpunct}
{\mcitedefaultendpunct}{\mcitedefaultseppunct}\relax
\EndOfBibitem
\bibitem[Hradil(1997)]{Hradil1997}
Hradil,~Z. Quantum-state estimation. \emph{Phys. Rev. A} \textbf{1997},
  \emph{55}, R1561\relax
\mciteBstWouldAddEndPuncttrue
\mciteSetBstMidEndSepPunct{\mcitedefaultmidpunct}
{\mcitedefaultendpunct}{\mcitedefaultseppunct}\relax
\EndOfBibitem
\bibitem[James \latin{et~al.}(2001)James, Kwiat, Munro, and White]{James2001}
James,~D. F.~V.; Kwiat,~P.~G.; Munro,~W.~J.; White,~A.~G. Measurement of
  qubits. \emph{Phys. Rev. A} \textbf{2001}, \emph{64}, 052312\relax
\mciteBstWouldAddEndPuncttrue
\mciteSetBstMidEndSepPunct{\mcitedefaultmidpunct}
{\mcitedefaultendpunct}{\mcitedefaultseppunct}\relax
\EndOfBibitem
\bibitem[Paris and {\v{R}}eh{\'a}{\v{c}}ek(2004)Paris, and
  {\v{R}}eh{\'a}{\v{c}}ek]{Paris2004}
Paris,~M. G.~A., {\v{R}}eh{\'a}{\v{c}}ek,~J., Eds. \emph{Quantum State
  Estimation}; Lecture Notes in Physics; Springer: Berlin, Heidelberg, 2004;
  Vol. 649\relax
\mciteBstWouldAddEndPuncttrue
\mciteSetBstMidEndSepPunct{\mcitedefaultmidpunct}
{\mcitedefaultendpunct}{\mcitedefaultseppunct}\relax
\EndOfBibitem
\bibitem[Verteletskyi \latin{et~al.}(2020)Verteletskyi, Yen, and
  Izmaylov]{Verteletskyi2020}
Verteletskyi,~V.; Yen,~T.-C.; Izmaylov,~A.~F. Measurement optimization in the
  variational quantum eigensolver using a minimum clique cover. \emph{J. Chem.
  Phys.} \textbf{2020}, \emph{152}, 124114\relax
\mciteBstWouldAddEndPuncttrue
\mciteSetBstMidEndSepPunct{\mcitedefaultmidpunct}
{\mcitedefaultendpunct}{\mcitedefaultseppunct}\relax
\EndOfBibitem
\bibitem[DalFavero \latin{et~al.}(2025)DalFavero, Sarkar, Rowland, Camps,
  Sawaya, and LaRose]{DalFavero2025}
DalFavero,~B.; Sarkar,~R.; Rowland,~J.; Camps,~D.; Sawaya,~N. P.~D.; LaRose,~R.
  Measurement reduction for expectation values via fine-grained commutativity.
  \emph{Phys. Rev. A} \textbf{2025}, \emph{112}, 052407\relax
\mciteBstWouldAddEndPuncttrue
\mciteSetBstMidEndSepPunct{\mcitedefaultmidpunct}
{\mcitedefaultendpunct}{\mcitedefaultseppunct}\relax
\EndOfBibitem
\bibitem[Nielsen and Chuang(2000)Nielsen, and Chuang]{Nielsen2000}
Nielsen,~M.~A.; Chuang,~I.~L. \emph{Quantum Computation and Quantum
  Information}; Cambridge University Press, 2000\relax
\mciteBstWouldAddEndPuncttrue
\mciteSetBstMidEndSepPunct{\mcitedefaultmidpunct}
{\mcitedefaultendpunct}{\mcitedefaultseppunct}\relax
\EndOfBibitem
\end{mcitethebibliography}

\end{document}